**Manipulation of Spin Transport in Graphene by Surface Chemical Doping**

K. Pi, Wei Han, K. M. McCreary, A. G. Swartz, Yan Li, and R. K. Kawakami[‡]

Department of Physics and Astronomy, University of California, Riverside, CA 92521

[‡]e-mail: roland.kawakami@ucr.edu

**Abstract**

The effects of surface chemical doping on spin transport in graphene are investigated by performing non-local measurements in ultrahigh vacuum while depositing gold adsorbates. We demonstrate manipulation of the gate-dependent non-local spin signal as a function of gold coverage. We discover that charged impurity scattering is not the dominant mechanism for spin relaxation in graphene, despite its importance for momentum scattering. Finally, unexpected enhancements of the spin lifetime illustrate the complex nature of spin relaxation in graphene.

PACS: 85.75.-d, 72.25.Rb, 72.25.Dc, 72.80.Vp

Graphene is an attractive material for spintronics [1] due to the observation of gate-tunable spin transport at room temperature [2-4] and the potential for long spin lifetimes (μs regime) arising from low intrinsic spin-orbit and hyperfine couplings [5-8]. The extreme surface sensitivity of graphene [9] introduces the possibility of manipulating the spin transport properties by surface chemical doping. Furthermore, this enables new methods for investigating fundamental questions such as the origin of spin relaxation in graphene, which has become a central issue since experiments have realized spin lifetimes up to only a few hundred picoseconds [2, 8, 10-16]. Recent experiments report that the primary source of spin relaxation is momentum scattering [12]. Furthermore, the dominant source of momentum scattering is believed to be charged impurity (CI) scattering [17-21]. Therefore, it is expected that CI scattering is an important factor for spin relaxation. In this Letter we investigate, for the first time, the effects of surface chemical doping on spin transport in graphene. Specifically, non-local spin signals of single-layer graphene spin valves are monitored *in situ* as gold adsorbates are deposited in ultrahigh vacuum and at cryogenic temperatures, and several important results are obtained. First, we demonstrate manipulation of the gate-dependent non-local spin signal as a function of sub-monolayer gold coverage. Second, we discover that CI scattering is not the dominant mechanism for spin relaxation, despite its importance for momentum scattering. Third, unexpected enhancements of the spin lifetime illustrate the complex nature of spin relaxation in graphene and demonstrate the concept of spin-preserved chemical doping, which produces a three-fold enhancement of spin lifetime at fixed gate voltages. These results address both fundamental and applied issues that are important for the development of graphene spintronics.

To investigate the role of CI scattering on spin relaxation, our approach is to systematically introduce additional sources of CI scattering and monitor their effect on spin lifetime. Gold impurities are selected for this purpose because they have been shown to behave as CI scatterers with 1/*r* Coulomb potential [19, 22] and are not expected to generate other effects such as resonant scattering, wavefunction hybridization, or chemical bonding [23-25]. It is known that CI scattering will generate spin relaxation, but the question is how this spin relaxation rate ($1/\tau_{CI}$) compares with the rates of other possible mechanisms (e.g. local lattice deformation,[16] substrate phonons,[15] etc.). If the various spin relaxation mechanisms are uncorrelated, then the total spin lifetime, $\tau_s$, is given by $1/\tau_s = 1/\tau_{CI} + \sum_j 1/\tau_j$, where the summation is over the other possible spin relaxation mechanisms. If CI scattering is the dominant spin relaxation rate, then $1/\tau_s \approx 1/\tau_{CI}$ and an increase in the amount of CI scattering will increase the overall spin relaxation rate. On the other hand, if another mechanism provides the dominant spin relaxation rate, then any changes to $1/\tau_{CI}$ will have negligible effect on the overall spin relaxation rate. Therefore, this study allows us to answer the critical question: Is charged impurity scattering important for spin relaxation in graphene?

We begin by measuring spin transport through graphene spin valves. Figure 1a is a schematic drawing of the graphene spin valve and the non-local measurement geometry. Single-layer graphene is obtained by mechanical exfoliation of HOPG onto $SiO_2$/Si substrate (300 nm thickness of $SiO_2$) and ferromagnetic electrodes with transparent contact are defined by electron beam lithography and angle evaporation of a 2 nm MgO masking layer followed by 80 nm Co [4, 11]. Spin transport measurements are performed inside a molecular beam epitaxy (MBE) chamber with a base pressure of $1\times10^{-10}$ torr. During the entire measurement, the sample is kept at low temperature (18 K) inside the chamber. For the non-local measurement, we apply a current (*I*) across electrodes E2 and E1 while measuring the voltage (*V*) between electrodes E3 and E4. When spin transport is present, the non-local resistance ($R_{nl} = V/I$) is positive for parallel alignment and negative for antiparallel alignment of the E2 and E3 magnetizations. Figure 1b shows non-local magnetoresistance for a device with electrode spacing (E2 to E3) of $L = 2.5$ μm and the gate voltage ($V_g$) tuned to the charge neutrality point ($V_D = -15$ V). The spin signal, $\Delta R_{nl}$, is defined by the difference between the parallel and antiparallel states, and has a value of ~580 mΩ. All data presented in this paper are measured from this sample, but we note that out of four devices studied (with electrode spacing from 1 – 4 μm), consistent results are obtained on all devices.

We examine the effect of Au doping on charge transport properties by employing *in situ* transport measurements combined with MBE growth in ultrahigh vacuum [22, 26]. The Au is deposited from a thermal effusion cell with a growth rate of 0.04 Å/min ($5\times10^{11}$ atom/cm$^2$s) while samples are maintained at 18 K to reduce the surface diffusion of the Au atoms [22]. For typical deposition times (less than 10 s total), the amount of Au is less than a percent of a monolayer. Figure 2a shows the gate-dependent conductivity (σ) for different amounts of Au coverage indicated by the deposition time (0 s, 2 s, 4 s, 6 s, 8 s) on a spin valve device. With increasing coverage, the Dirac point shifts toward negative gate voltage, which indicates that Au donates electrons to the graphene and the Au impurities become positively charged. Electron and hole mobilities are obtained by measuring the slope of the conductivity curve away from the Dirac point. Figure 2c shows that mobilities for electrons and holes decrease as a function of Au coverage. Despite the small amount of Au deposited, the mobility diminishes by more than half. This indicates that CI scattering introduced by Au impurities exceeds the scattering initially present in the clean device.

We next consider the effect of the Au doping on the spin transport properties. A model for lateral spin transport was developed by Takahashi *et. al.* [27], and we apply the same method to obtain an expression for the non-local spin signal of graphene spin valves with transparent contacts:

$$\Delta R_{nl} = C\sigma f(\lambda_s), \quad C = \frac{4P_F^2}{(1-P_F^2)^2}\frac{\rho_F^2 \lambda_F^2}{A_J^2}\frac{W}{L}, \quad f(\lambda_s) = \frac{L}{\lambda_s}\left[\frac{\exp(-L/\lambda_s)}{1-\exp(-2L/\lambda_s)}\right] \quad (1)$$

where $L$ is the electrode spacing, $W$ is the graphene width, $\lambda_s$ is the spin diffusion length of graphene, $P_F$ is the spin polarization of the ferromagnet, and $\rho_F$ is the resistivity of the ferromagnet, $\lambda_F$ is the spin diffusion length of the ferromagnet, and $A_J$ is the area of the junction between the ferromagnet and graphene. The prefactor, $C$, consists of terms that are not altered by Au doping, and therefore we can treat this as a constant factor for all $\Delta R_{nl}$ data. Doping can affect $\Delta R_{nl}$ either through changes in the spin diffusion length or the (spin independent) conductivity. Figure 2b shows the gate dependence of the non-local magnetoresistance for different Au coverages (0 s, 2 s, 4 s, 6 s, 8 s). In agreement with equation (1), $\Delta R_{nl}$ tracks the gate dependence of the conductivity, with minimum values of the $\Delta R_{nl}$ curves coinciding with the Dirac point. Furthermore, the overall magnitude of $\Delta R_{nl}$ remains relatively unchanged, which indicates that the Au atoms do not dramatically suppress the spin diffusion length. This demonstrates that the manipulation of the non-local spin signal is mostly attributed to the effect of Au doping on the conductivity.

To quantify the effect that Au doping does have on the spin diffusion length, we consider the ratio $\Delta R_{nl}/\sigma$. Referring to equation (1), this ratio factors out the strong effect that the Au doping has on the conductivity, thus leaving only the part that depends on the spin diffusion length [ $\Delta R_{nl}/\sigma = C f(\lambda_s)$ ]. Because $f$ is a monotonic function of $\lambda_s$ (see Figure 2d inset, with $L = 2.5$ μm), an increase (decrease) in $\Delta R_{nl}/\sigma$ corresponds to an increase (decrease) in $\lambda_s$. At typical values of $\lambda_s$ between 1 and 3 μm, the value of $f$ varies significantly. For each Au coverage, we average the value of $\Delta R_{nl}/\sigma$ over gate voltages within ±35 V of the Dirac point. As shown in Figure 2d, $\langle \Delta R_{nl}/\sigma \rangle$ does not decrease with Au coverage, which implies that the average $\lambda_s$ does not decrease with Au coverage. This behavior shows that CI scattering is not important for spin relaxation in graphene. An unexpected trend is that $\langle \Delta R_{nl}/\sigma \rangle$, and hence $\lambda_s$, actually increases slightly with Au coverage. This suggests that the Au impurities produce a slight suppression of the spin relaxation.

The role of CI scattering on spin relaxation is investigated further with a different type of experiment. We directly measure $\tau_s$ by applying an out-of-plane magnetic field to induce electron spin precession (Hanle effect) as the spins diffuse from E2 to E3 in the non-local geometry [2, 11]. Figure 3a and 3b show the representative Hanle data and best fit curves for different amounts of Au coverage (0 s and 8 s). The top (bottom) curve is the non-local resistance for the parallel (antiparallel) alignment of the E2 and E3 magnetizations. For the fitting, we first obtain the symmetric part of the data, $[R_{nl}(H)+R_{nl}(-H)]/2$, to remove effects unrelated to spin. A constant background is subtracted and the parallel and antiparallel data are simultaneously fit with the following equation using the same values for $\tau_s$ and $D$, but different values for the amplitude $S_0$:

$$R_{nl}^{P(AP)} = (\pm)S_0 \int \frac{1}{\sqrt{4\pi Dt}} \exp\left[-\frac{L^2}{4Dt}\right] \cos(g\mu_B H_\perp/\hbar) \exp(-t/\tau_s) dt \qquad (2)$$

where $g$ is the g-factor, $\mu_B$ is the Bohr magneton, $H_\perp$ is the out-of-plane magnetic field, the electrode spacing $L$ is 2.5 µm, and the + (-) sign is for the parallel (antiparallel) magnetization state. Figure 3c shows the best fit values of $\tau_s$ as a function of Au coverage at the Dirac point (black squares), for an electron concentration of $2.9\times10^{12}$ cm$^{-2}$ (red circles), and for a hole concentration of $2.9\times10^{12}$ cm$^{-2}$ (blue triangles). In all three cases, $\tau_s$ shows a slight enhancement with increasing Au coverage. The inset of Figure 3c shows the corresponding values of *D*, which decrease as a function of Au coverage.

The relationship between $\tau_s$ and $\tau_m$, the momentum scattering time, is particularly important for understanding the spin relaxation. There are two mechanisms by which spin-orbit coupling could generate spin relaxation in graphene [28]. One is the Elliot-Yafet (EY) mechanism, in which a spin flip occurs with finite probability during a momentum scattering event, resulting in $\tau_s \sim \tau_m$. The other is the Dyakonov-Perel (DP) mechanism, where spins precess in internal spin-orbit fields, with $\tau_s \sim \tau_m^{-1}$. Recent work by van Wees shows that $\tau_s \sim D$ ($\sim \tau_m$, specifically $D = (1/2)v_F^2\tau_m$ where $v_F$ is the Fermi velocity [12]) for graphene spin valves, indicating that the EY mechanism is more important than the DP mechanism [12]. On the other hand, in Figure 3c we observe that *D* decreases with Au coverage while $\tau_s$ exhibits a slight increase, illustrating that the $\tau_s \sim D$ scaling is not obeyed. This does not necessarily argue against EY, but rather points out that many different types of momentum scattering can produce EY spin relaxation (i.e. CI scattering, phonons, local sp$^3$ deformation, edge scattering, short-range impurity potential, etc.) [14-16], and each type of momentum scattering can have a different efficiency for spin scattering. Stated in another way, if the dominant momentum scattering mechanism is different from the dominant EY spin relaxation mechanism, then $\tau_m$ and $\tau_s$ need not show any particular relation. Our results indicate that CI scattering is very effective for momentum scattering, but it produces negligible spin relaxation on the time scale of ~100 ps.

While the lack of a strong decrease of $\tau_s$ establishes that CI scattering is not the primary source of spin relaxation, the slight increase of $\tau_s$ is rather unexpected and brings up some interesting issues. There are several possible effects which may result in the slight increase of $\tau_s$ with Au coverage. The first possibility is that the spin relaxation mechanisms are correlated. For uncorrelated mechanisms, the spin relaxation rates simply add (as discussed earlier) so that any additional relaxation cannot increase the overall spin lifetime. In our case, a correlation might arise if the gold impurities actively inhibit some of the other spin relaxation mechanisms. For instance, the gold may bind to lattice defects or edge sites [29, 30], reducing their ability to scatter spins. A second possibility is that DP makes a non-negligible contribution to the overall spin relaxation. While the scaling of $\tau_s$ with $D$ observed by van Wees [12] favors the EY-type mechanisms and the intrinsic DP mechanism should be very weak due to the low intrinsic spin-orbit coupling [5-7], some theoretical calculations suggest that a curvature-enhanced DP mechanism could contribute to the overall spin lifetime [5]. If DP provides an appreciable contribution, then the slight enhancement could occur because $\tau_S^{DP} \sim \tau_m^{-1}$. The unanticipated increase of $\tau_s$ highlights the complex nature of spin relaxation in graphene and motivates further systematic studies of its origin.

The robustness of spin polarization against CI scattering may prove to be useful for future applications. In principle, impurities could be used to tune the carrier concentration without inducing additional spin relaxation. To demonstrate this idea, we investigate $\Delta R_{nl}$ and $\sigma$ as a function of Au doping at fixed gate voltage ($V_g$ = -15 V, the initial Dirac point). As shown in Figure 4a, $\sigma$ increases as a result of the substantial increase in carrier concentration, and $\Delta R_{nl}$ follows the trend due to the linear scaling with $\sigma$ (equation 1). Figure 4b shows a significant enhancement in the spin lifetime, $\tau_s$, from 50 ps to 150 ps at the highest Au coverage. This is a fortuitous situation where $\tau_s$ increases due to the higher carrier concentration [12] combined with the lack of additional spin relaxation. This demonstrates the concept of spin-preserved chemical doping of carriers, which provides a potentially useful tool in the design of future graphene spintronic devices.

In conclusion, we have demonstrated the manipulation of spin transport properties in graphene through chemical doping. The results clearly show that CI scattering is not the primary source of spin relaxation in graphene, even though it is very effective at generating momentum scattering. Additionally, an enhancement of spin lifetime is observed with increased gold coverage, which may become useful for future graphene spintronic devices.

We acknowledge stimulating discussions with E. Johnston-Halperin and acknowledge the support of ONR (N00014-09-1-0117), NSF (CAREER DMR-0450037), and NSF (MRSEC DMR-0820414).

**FIGURE CAPTIONS**

**Figure 1.** (a) Schematic drawing of the graphene spin valve and the non-local measurement geometry. (b) Non-local magnetoresistance curves measured at 18 K with gate voltage tuned to the Dirac point (-15 V). The red (black) curve is for increasing (decreasing) magnetic field.

**Figure 2.** (a) Gate dependent conductivity for Au deposition times of 0 s, 2 s, 4 s, 6 s and 8 s at a deposition rate of 0.04 Å/min. (b) Gate dependence of non-local spin signal for Au deposition times of 0 s, 2 s, 4 s, 6 s and 8 s. (c) Mobility for electrons (red open circles) and holes (blue solid circles) for different Au coverages. (d) $\langle \Delta R_{nl}/\sigma \rangle$ for different Au coverages. Inset: $f$ vs. spin diffusion length for $L$=2.5 μm.

**Figure 3.** (a, b) Hanle spin precession data (open circles) and the best fit curves (red line) at hole concentration of $2.9 \times 10^{12}$ cm$^{-2}$ for 0 s and 8 s of Au deposition. (c) Dependence of spin lifetime on Au coverage for electrons with concentration of $2.9 \times 10^{12}$ cm$^{-2}$ (red circles), at the charge neutrality point (blue triangles), and for holes with concentration $2.9 \times 10^{12}$ cm$^{-2}$ (black squares). Error bars represent the 99% confidence interval. Inset: the diffusion constant vs. Au deposition time. Error bars are omitted when they are comparable to the symbol size.

**Figure 4.** (a) Dependence of $\Delta R_{nl}$ (black squares, left axis) and conductivity (red circles, right axis) on Au coverage with gate voltage held constant at the initial Dirac point ($V_g$ = -15 V). (b) Spin lifetime as a function of Au deposition time. Error bars represent the 99% confidence interval. Inset: Diffusion constant as a function of Au deposition time. The weak variation of $D$ with Au coverage occurs because the increased scattering reduces $D$, while the increased carrier concentration increases $D$. Error bars are omitted when they are comparable to the symbol size.

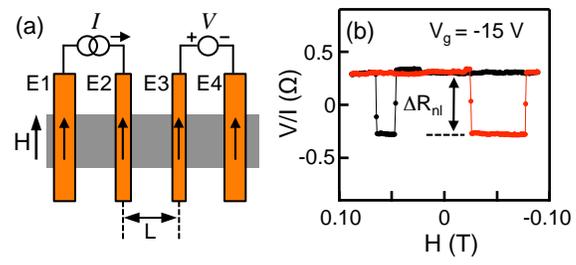

Figure 1.

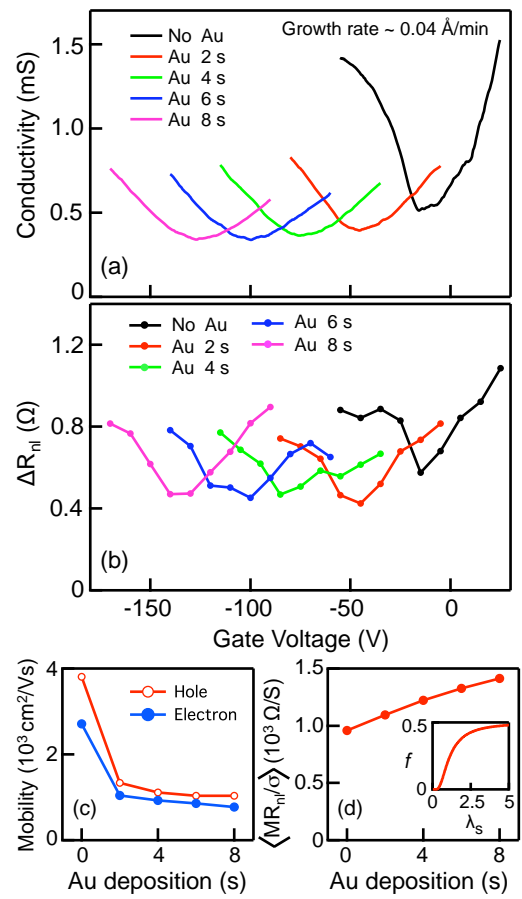

Figure 2.

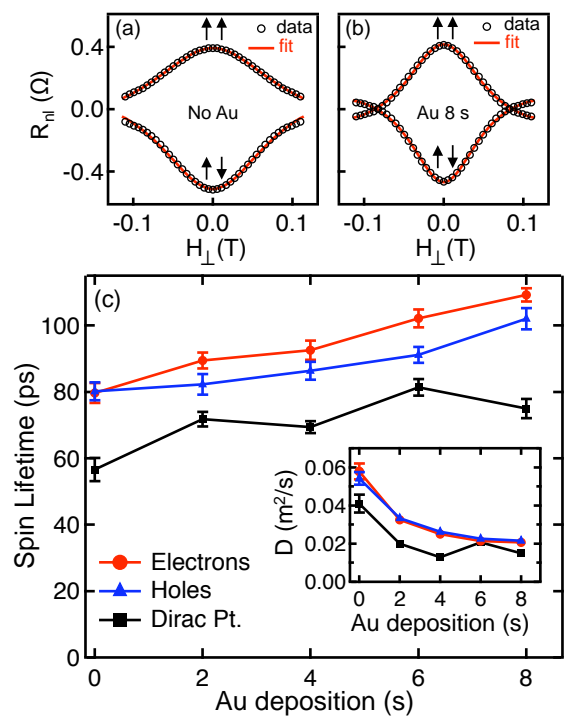

Figure 3.

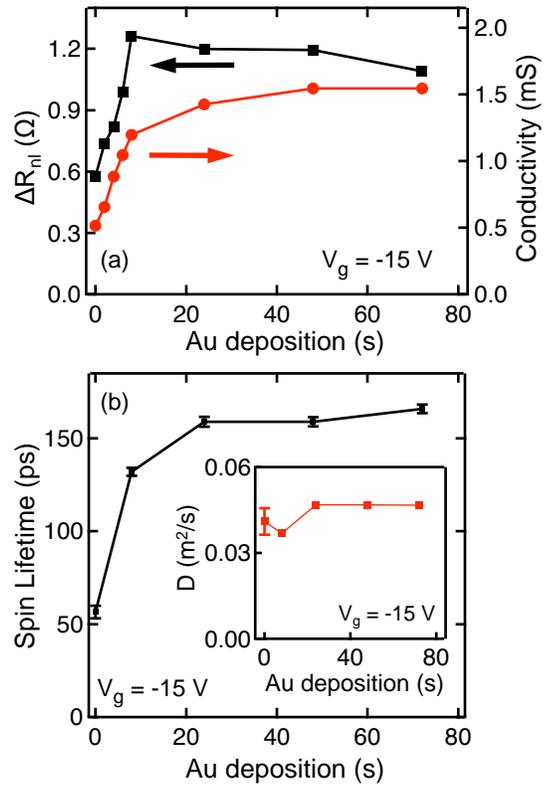

Figure 4.